\documentclass{article}
\usepackage{spconf,amsmath,graphicx,amsfonts,amssymb,stfloats,booktabs,multirow,bbding,pifont,url,color,xcolor,makecell,bm,cite,diagbox}
\usepackage[colorlinks,linkcolor=black,anchorcolor=black,citecolor=black,urlcolor=black]{hyperref}

\def\L{{\cal L}}

\title{Gradient Remedy for Multi-Task Learning in End-to-End Noise-Robust Speech Recognition}
\name{Yuchen Hu$^1$, Chen Chen$^1$, Ruizhe Li$^2$, Qiushi Zhu$^3$, Eng Siong Chng$^1$}
\address{$^1$Nanyang Technological University, Singapore \quad $^2$University of Aberdeen, UK\\ $^3$University of Science and Technology of China, China}

\begin{document}
\ninept
\maketitle
\begin{abstract}
Speech enhancement (SE) is proved effective in reducing noise from noisy speech signals for downstream automatic speech recognition (ASR), where multi-task learning strategy is employed to jointly optimize these two tasks.
However, the enhanced speech learned by SE objective may not always yield good ASR results.
From the optimization view, there sometimes exists interference between the gradients of SE and ASR tasks, which could hinder the multi-task learning and finally lead to sub-optimal ASR performance.
In this paper, we propose a simple yet effective approach called gradient remedy (GR) to solve interference between task gradients in noise-robust speech recognition, from perspectives of both angle and magnitude.
Specifically, we first project the SE task's gradient onto a dynamic surface that is at acute angle to ASR gradient, in order to remove the conflict between them and assist in ASR optimization.
Furthermore, we adaptively rescale the magnitude of two gradients to prevent the dominant ASR task from being misled by SE gradient.
Experimental results show that the proposed approach well resolves the gradient interference and achieves relative word error rate (WER) reductions of 9.3\% and 11.1\% over multi-task learning baseline, on RATS and CHiME-4 datasets, respectively.
Our code is available at GitHub\footnote{\url{https://github.com/YUCHEN005/Gradient-Remedy}}.
\end{abstract}



\vspace{-0.08cm}
\section{Introduction}
\label{sec:intro}
\begin{figure}[t!]
  \centering
  \includegraphics[width=1.0\columnwidth]{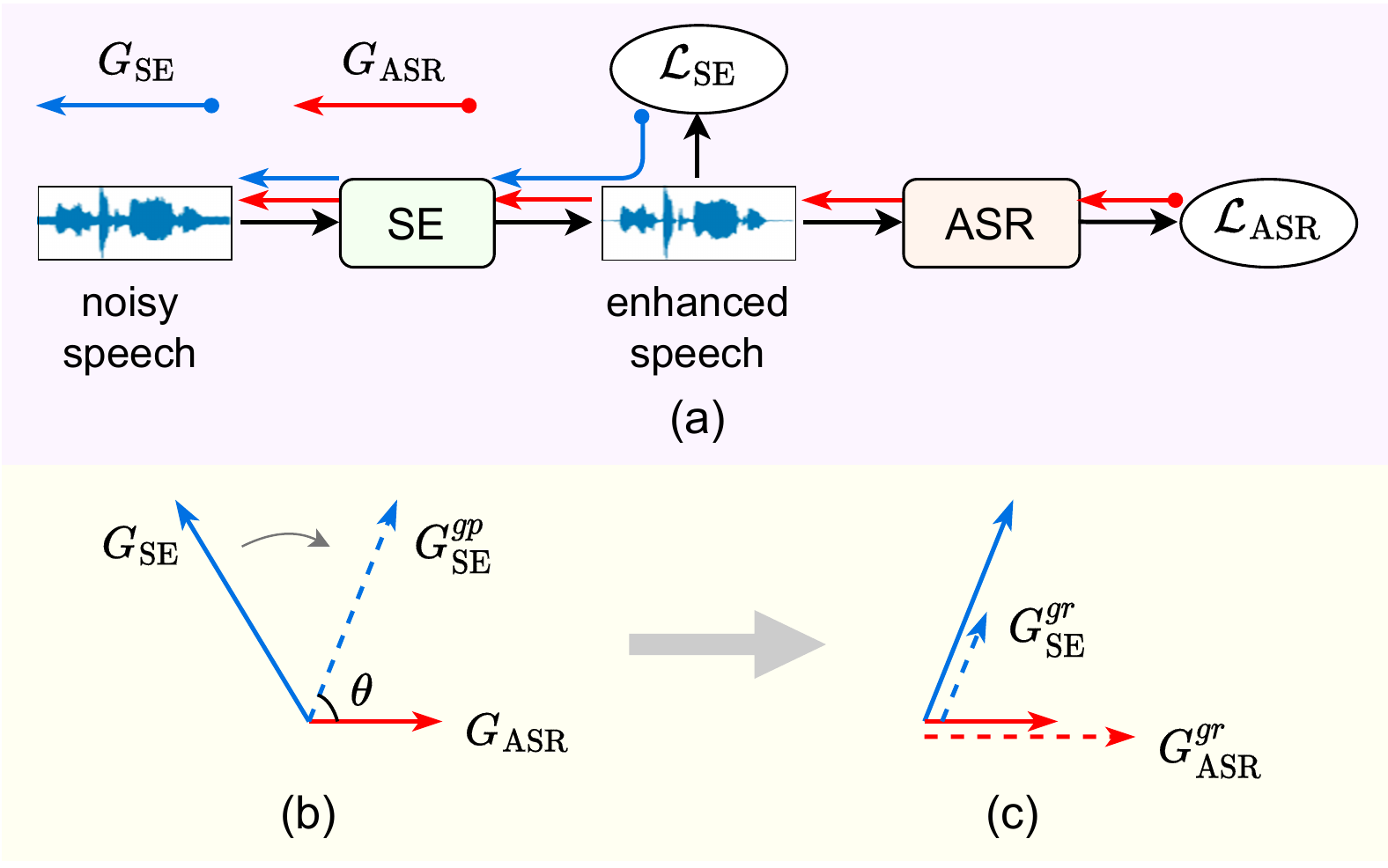}
  \vspace{-0.6cm}
  \caption{Block diagrams of (a) multi-task learning of SE and ASR, and our gradient remedy approach: (b) gradient projection, (c) gradient rescale. 
  The $\bm{G}$ denotes gradient, and $\theta$ is a dynamic acute angle.}\label{fig1}
  \vspace{-0.5cm}
\end{figure}

\vspace{-0.15cm}
Speech enhancement (SE)~\cite{Wang2014on, pascual2017segan, wang2020complex,zheng2021interactive} is proved effective in reducing noise from the noisy speech signals to improve speech quality for downstream tasks, \textit{e.g.}, automatic speech recognition (ASR)~\cite{chen2022noise,chen2022self,zhu2022noise,zhu2022joint,zhu2022robust}.
Prior work~\cite{subramanian2019speech} proposed a cascaded SE and ASR system using final ASR training objective for optimization.
Later studies~\cite{pandey2021dual, ma2021multitask} believed that the SE training objective can direct the enhancement module to produce better enhanced speech for downstream ASR.
Therefore, they proposed a multi-task learning strategy to jointly optimize the SE and ASR tasks, as shown in Figure~\ref{fig1}(a).
In this way, the front-end SE module is supervised by both tasks, where ASR is the dominant task we target at and SE serves as an auxiliary task to benefit ASR.

However, recent work~\cite{hu2022interactive,hu2022dual} found that apart from noise, SE could also reduce some speech information important for ASR, so that the enhanced speech learned by SE objective may not always yield good ASR results.
From the optimization view, we can observe some interference between the gradients of SE and ASR tasks, which could hinder the multi-task learning and finally degrade the ASR performance.
Firstly, the angle between two gradients sometimes exceeds $90^{\circ}$ (See Figure~\ref{fig1}(b)), which means that they are conflicting.
Therefore, the auxiliary SE task would hinder, instead of assist in, the optimization of dominant ASR task.
Secondly, the magnitude of SE gradient sometimes becomes much larger than ASR gradient (See solid arrows in Figure~\ref{fig1}(c)), which we define as wrongly dominant since SE is only the auxiliary task.
In this case, it would dominate the overall gradient and thus mislead the ASR task's optimization.

Recent work~\cite{yu2020gradient,zhu2022prompt} proposed a projecting conflicting gradients (PCGrad) method to avoid conflict in multi-task learning, which projects a task's gradient onto the normal plane of other conflicting task gradients, as shown in Figure~\ref{fig2}(a).
In this way, it could remove the conflict but may not solve the problem of wrongly dominant gradient.
Some other works~\cite{guo2018dynamic,groenendijk2021multi} proposed to learn dynamic weights to balance different training objectives in multi-task learning, from the perspective of loss optimization trends, which could be effective in alleviating the wrongly dominant SE gradient.
However, they did not consider the priority of different tasks, so that this strategy may also weaken the dominant role of ASR in our system.

In this paper, we propose a simple yet effective approach called gradient remedy (GR) to solve interference between task gradients in noise-robust speech recognition, from perspectives of both angle and magnitude.
Specifically, we first project the conflicting SE gradient onto a dynamic surface that is at acute angle to ASR gradient, instead of the normal plane in PCGrad.
In this way, we can not only remove the conflict between them, but also push SE gradient to help optimize the dominant ASR task.
Furthermore, in case of wrongly dominant SE gradient, we adaptively rescale the magnitude of two gradients to prevent it from misleading the ASR task, highlighting the dominant role of ASR in our system.
Experimental results show that our GR approach well resolves the gradient interference and improves the final ASR performance under different-level noisy conditions.
To the best of our knowledge, this is the first work to harmonize gradients for multi-task learning in noise-robust speech recognition.

\vspace{-0.2cm}
\section{Proposed Method}
\label{sec:proposed_method}
\vspace{-0.2cm}
\subsection{System Overview}
\label{ssec:system_overview}
\vspace{-0.1cm}
As illustrated in Figure~\ref{fig1}(a), we follow the architecture of multi-task learning system.
The noisy speech is first sent into speech enhancement module to obtain the enhanced speech, so that we can calculate a SE loss $\mathcal{L}_\text{SE}$ by comparing it with the ground-truth clean speech.
Then, the enhanced speech is sent into the ASR module to generate recognized tokens, which are used to calculate the cross-entropy based ASR loss $\mathcal{L}_\text{ASR}$ by compared to the ground-truth transcriptions.
According to the multi-task learning strategy, these two losses would be weight summed to form the final training objective: $\mathcal{L} = (1 - \lambda_\text{ASR}) \cdot \mathcal{L}_\text{SE} + \lambda_\text{ASR} \cdot \mathcal{L}_\text{ASR}$, where $\lambda_\text{ASR}$ is a weighting parameter to balance two objectives.

From the back-propagation view, we denote the SE task gradient as $\bm{G}_\text{SE} = \nabla_v [(1 - \lambda_\text{ASR})\cdot\mathcal{L}_\text{SE}]$, and the ASR task gradient as $\bm{G}_\text{ASR} = \nabla_v [\lambda_\text{ASR}\cdot\mathcal{L}_\text{ASR}]$, where $v$ stands for model parameters.
As shown in Figure~\ref{fig1}, $\bm{G}_\text{SE}$ goes back through the SE module only (blue arrows), and $\bm{G}_\text{ASR}$ passes both ASR and SE modules (red arrows).
Therefore, the SE module would be optimized by both gradients, thus its overall gradient can be expressed as:
\vspace{-0.2cm}
\begin{equation}
\label{eq1}
\begin{split}
    \bm{G} &= \bm{G}_\text{SE} + \bm{G}_\text{ASR},
\end{split}
\end{equation}

\vspace{-0.2cm}
However, there sometimes exists interference between the task gradients $\bm{G}_\text{SE}$ and $\bm{G}_\text{ASR}$, which could hinder the multi-task learning and finally lead to sub-optimal ASR performance.
To this end, we propose a gradient remedy approach to solve the interference problem, which generates two new task gradients that harmony with each other, \textit{i.e.}, $\bm{G}^{gr}_\text{SE}$ and $\bm{G}^{gr}_\text{ASR}$.
Then we obtain the final gradient $\bm{G}^{gr}$ as follows for SE module's optimization:
\vspace{-0.2cm}
\begin{equation}
\label{eq2}
\begin{split}
    \bm{G}^{gr} &= \bm{G}^{gr}_\text{SE} + \bm{G}^{gr}_\text{ASR},
\end{split}
\end{equation}

\begin{figure}[t]
  \centering
  \vspace{-0.05cm}
  \includegraphics[width=0.8\columnwidth]{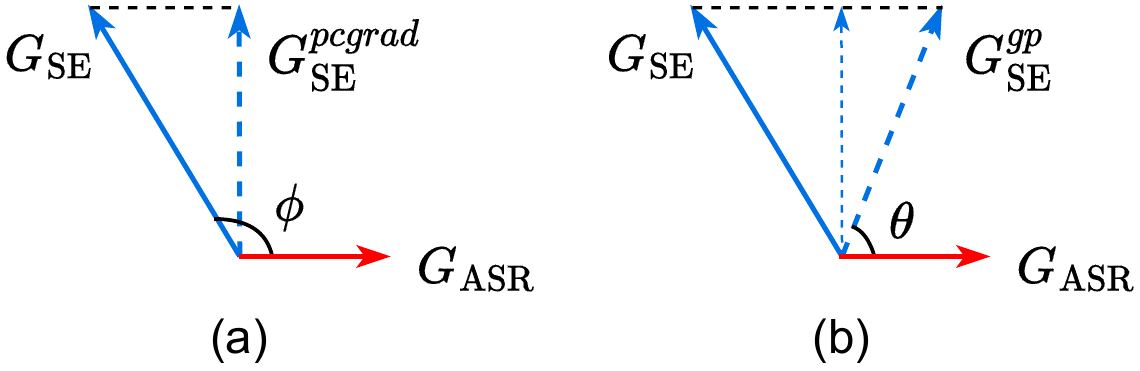}
  \vspace{-0.3cm}
  \caption{Comparison of (a) PCGrad method~\cite{yu2020gradient}, and (b) our proposed gradient projection method.}\label{fig2}
  \vspace{-0.35cm}
\end{figure}

\vspace{-0.5cm}
\subsection{Interfered Gradients}
\label{ssec:interfered_grad}
\vspace{-0.1cm}
During the multi-task learning process, we can observe some interference between the gradients of SE and ASR tasks, indicating that the auxiliary SE task may not always benefit the dominant ASR task.
Such interference can be classified into two categories, \textit{i.e.}, conflicting gradient and wrongly dominant gradient.

\vspace{0.1cm}
\noindent\textbf{Conflicting Gradient.} We denote the angle between task gradients $\bm{G}_\text{SE}$ and $\bm{G}_\text{ASR}$ as $\phi$, and we define the gradients as conflicting when $\phi > 90^\circ$.
As shown in Figure~\ref{fig2}, $\bm{G}_\text{SE}$ and $\bm{G}_\text{ASR}$ are contradicting each other, which means the auxiliary SE task is hindering, instead of assisting in, the dominant ASR task.

\vspace{0.1cm}
\noindent\textbf{Wrongly Dominant Gradient.} We define SE gradient as wrongly dominant when its magnitude is much larger than that of ASR gradient (See Figure~\ref{fig3}), \textit{i.e.}, $\Vert \bm{G}^{gp}_\text{SE} \Vert_2 > K \cdot \Vert \bm{G}_\text{ASR} \Vert_2$, where $K > 1$ is a threshold hyper-parameter.
In this case, the overall gradient that equals to their sum would be dominated by the SE gradient, which would mislead the optimization of dominant ASR task and finally degrade its performance.
In particular, the larger angle between them (denoted as $\theta'$ in Figure~\ref{fig3}), the overall gradient will deviate more from ASR gradient, which results in more misleading.

\subsection{Gradient Remedy (GR)}
\label{ssec:grad_remedy}
\vspace{-0.1cm}
In this work, we propose a simple yet effective approach called gradient remedy to solve interference between SE and ASR gradients.
Specifically, we first propose a gradient projection method to remove the conflicting component in SE gradient, as well as push it to assist in ASR optimization.
Furthermore, we design a gradient rescale strategy to prevent the dominant
ASR task from being misled by SE gradient, highlighting the dominant role of ASR in our system.

In particular, we operate on the gradients of each layer in SE module, which are flatten to 1-dimensional long vectors in advance and reshaped back after remedy to form the final gradient.

\vspace{-0.25cm}
\subsubsection{Gradient Projection}
\label{sssec:grad_projection}
\vspace{-0.1cm}
As shown in Figure~\ref{fig2}(a), PCGrad projects $\bm{G}_\text{SE}$ onto the normal plane of $\bm{G}_\text{ASR}$, removing the conflicting component in SE gradient.
Differ to PCGrad, we propose a novel method in Figure~\ref{fig2}(b) to project $\bm{G}_\text{SE}$ onto a dynamic surface that is at acute angle $\theta$ to $\bm{G}_\text{ASR}$.
In this way, we can not only remove the conflict, but also push SE gradient to help optimize the dominant ASR task.
According to Figure~\ref{fig2}(b), our gradient projection method can be mathematically formulated as:
\begin{equation}
\label{eq3}
\bm{G}^{gp}_\text{SE}=
\left\{
\begin{array}{ll}
    \hspace{-0.1cm}\bm{G}_\text{SE} + \Vert \bm{G}_\text{SE} \Vert_2 \cdot \left(\frac{\sin \phi}{\tan \theta} - \cos \phi\right) \cdot \frac{\bm{G}_\text{ASR}}{\Vert \bm{G}_\text{ASR} \Vert_2}, &\hspace{-0.1cm} \text{if}\hspace{0.13cm} \phi > 90^\circ \\
    \hspace{-0.1cm}\bm{G}_\text{SE}, &\hspace{-0.1cm} \text{otherwise}. 
\end{array}
\right.
\end{equation}

\noindent where $\phi$ is the angle between $\bm{G}_\text{SE}$ and $\bm{G}_\text{ASR}$, and $\theta$ is a dynamic angle that we design as:
\vspace{-0.2cm}
\begin{equation}
\label{eq4}
\begin{split}
    \theta &= \arctan \frac{\Vert \bm{G}_\text{SE} \Vert_2}{\Vert \bm{G}_\text{ASR} \Vert_2},
\end{split}
\end{equation}
\vspace{-0.3cm}

\noindent where $\frac{\Vert \bm{G}_\text{SE} \Vert_2}{\Vert \bm{G}_\text{ASR} \Vert_2} \in (0, +\infty)$, so that $\theta \in (0, 90^\circ)$.
The idea behind this design is to control steady projection.
In particular, when the magnitude of SE gradient is small relative to ASR gradient, we set a small $\theta$ to push $\bm{G}_\text{SE}$ more towards the $\bm{G}_\text{ASR}$, \textit{i.e.}, large $\frac{\sin \phi}{\tan \theta}$ in Equation~\ref{eq3}.
On the contrary, if the magnitude of $\bm{G}_\text{SE}$ is much larger than $\bm{G}_\text{ASR}$, then we employ a relatively large $\theta$ to project $\bm{G}_\text{SE}$ less towards the direction of $\bm{G}_\text{ASR}$.
Therefore, with such a dynamic $\theta$, we can control the projected SE gradient $\bm{G}^{gp}_\text{SE}$ to steadily assist in ASR optimization, which stabilizes the entire system training.

According to Equation~\ref{eq3}, the resulted angle between $\bm{G}^{gp}_\text{SE}$ and $\bm{G}_\text{ASR}$ equals to either $\theta$ or $\phi$, depending on whether $\phi$ exceeds $90^\circ$.
For simplicity, we denote it as $\theta'$ in Figure~\ref{fig3} and following sections.

\vspace{-0.25cm}
\subsubsection{Gradient Rescale}
\label{sssec:grad_rescale}
\vspace{-0.1cm}

\begin{figure}[t]
  \centering
  \includegraphics[width=0.74\columnwidth]{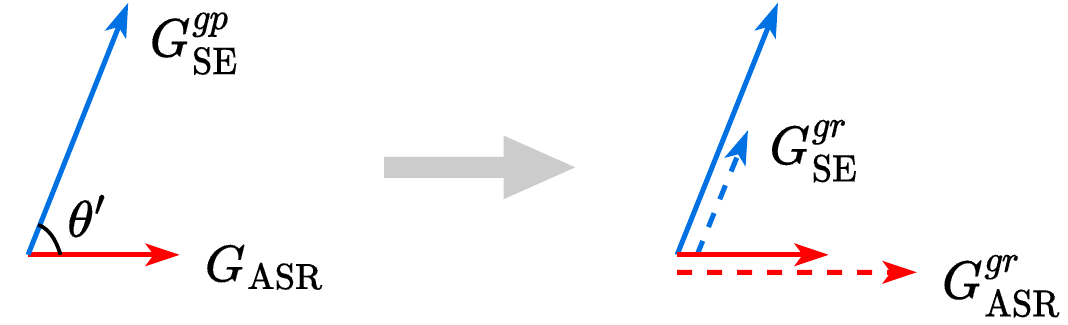}
  \vspace{-0.4cm}
  \caption{Block diagram of our proposed gradient rescale strategy.}\label{fig3}
  \vspace{-0.3cm}
\end{figure}

After gradient projection, the conflicting gradient can be avoided but the wrongly dominant SE gradient may still exist.
To further solve this problem, we propose a gradient rescale strategy in Figure~\ref{fig3} to adaptively compress the SE gradient and stretch the ASR gradient, in case that $\Vert \bm{G}^{gp}_\text{SE} \Vert_2 > K \cdot \Vert \bm{G}_\text{ASR} \Vert_2$, where $K$ is the threshold.
Our strategy is mathematically formulated as:
\begin{equation}
\label{eq5}
\left(\bm{G}^{gr}_\text{SE}, \bm{G}^{gr}_\text{ASR}\right) =
\left\{
\begin{array}{ll}
    \hspace{-0.15cm}\left(\bm{G}^{gp}_\text{SE} \cdot r, \bm{G}_\text{ASR} \hspace{0.02cm}/\hspace{0.02cm} r\right), &\hspace{-0.2cm}\text{if}\hspace{0.13cm} \Vert \bm{G}^{gp}_\text{SE} \Vert_2 > K \cdot \Vert \bm{G}_\text{ASR} \Vert_2 \vspace{0.1cm}\\
    \hspace{-0.15cm}\left(\bm{G}^{gp}_\text{SE}, \bm{G}_\text{ASR}\right), &\hspace{-0.2cm}\text{otherwise}. 
\end{array}
\right.
\end{equation}

\noindent we set an adaptive ratio $r=\cos\theta'$ to compress the $\bm{G}^{gp}_\text{SE}$.
The idea behind it is that, in case $\theta'$ is small, which means $\bm{G}^{gp}_\text{SE}$ aligns well with $\bm{G}_\text{ASR}$, so that the overall gradient will not deviate much from $\bm{G}_\text{ASR}$, resulting in less misleading.
In this case, it is unnecessary to compress $\bm{G}^{gp}_\text{SE}$ a lot, so that we set a relatively large ratio $r = \cos\theta'$.
In contrast, when $\theta'$ increases, the wrongly dominant $\bm{G}^{gp}_\text{SE}$ would increasingly mislead the ASR optimization, thus we need a smaller $r$ to perform more compression on $\bm{G}^{gp}_\text{SE}$.
In this way, we can adaptively compress SE gradient to avoid misleading the dominant ASR task. 

Apart from compressing SE gradient, we also use $1/r$ as ratio to stretch the ASR gradient accordingly, in order to further highlight the dominant role of ASR in our system.

\vspace{-0.1cm}
\section{Experiments and Results}
\label{sec:exp_result}

\vspace{-0.2cm}
\subsection{Datasets}
\label{ssec:datasets}
\vspace{-0.1cm}
We conduct experiments on two datasets, one is Robust Automatic Transcription of Speech (RATS)~\cite{graff2014rats} which consists of extremely noisy radio communication speech, and the other is CHiME-4~\cite{vincent2016chime4} dataset that contains far-field speech under normal noisy conditions.

The RATS dataset comprises eight parallel channels and in this work we use the Channel-A subset only, which consists of 44 hours of training data, 5 hours of valid data and 8 hours of test data. 
Since RATS dataset is chargeable by LDC, we release its fbank features and several listening samples on GitHub for reference\footnote{\scriptsize\url{https://github.com/YUCHEN005/RATS-Channel-A-Speech-Data}}.

The CHiME-4 dataset\footnote{\scriptsize\url{https://spandh.dcs.shef.ac.uk/chime_challenge/CHiME4}} consists of three partitions: clean data, real noisy data and simulated noisy data. 
The clean data is based on WSJ0~\cite{paul1992design} training set (si\_tr\_s). 
The real noisy data is recorded in four different noisy environments, \textit{i.e.}, bus, cafe, pedestrian area and street junction. 
The simulated noisy data is generated by mixing the clean data with background noise recorded in the above four environments. 
In this work, we utilize both the real and simulated noisy data of 1-channel track to evaluate our method.

\vspace{-0.2cm}
\subsection{Experimental Setup}
\label{ssec:exp_setup}
\vspace{-0.08cm}
\subsubsection{Network Configurations}
\label{sssec:network_configs}
\vspace{-0.1cm}
The multi-task learning system consists of two modules: SE module and ASR module. 
The SE module follows prior work~\cite{ma2021multitask} to predict a mask for noisy speech feature's magnitude, using 3 layers of 896-unit bidirectional long short-term memory (BLSTM)~\cite{hochreiter1997lstm} and a 257-unit linear layer followed by ReLU activation function, which contain 43.7 millions of parameters. 
Then we leverage the state-of-the-art Conformer~\cite{gulati2020conformer} for ASR module, where its encoder consists of 12 Conformer blocks, and the decoder contains 6 transformer~\cite{vaswani2017attention} blocks, with the embedding dimension/feed-forward dimension/attention heads set to 256/2048/4.
We employ 1000 byte-pair-encoding (BPE)~\cite{kudo2018bpe} tokens to model the ASR output. 

The system is optimized by Adam algorithm~\cite{kingma2014adam}, where the learning rate warms up linearly to 0.002 in first 25,000 steps and then decreases proportional to the inverse square root of training steps.
We train 50 epochs for experiments on RATS dataset and 100 epochs for CHiME-4 dataset.
The weighting parameter $\lambda_\text{ASR}$ is set to 0.7, the threshold $K$ is set to 5, and batch size is set to 64.
We also build a 2-layer 650-unit RNNLM on training text for rescoring during inference.
All hyper-parameters are tuned on validation set.

\vspace{-0.3cm}
\subsubsection{Reference Baselines}
\label{sssec:baselines}
\vspace{-0.1cm}
We build five competitive baselines to evaluate our proposed GR approach.
For fair comparison, we adopt same architectures and configurations for all the SE modules and ASR modules included.
Therefore, our approach requires no extra model parameters compared to the multi-task learning baseline.

\begin{enumerate}
\item \textbf{E2E ASR}~\cite{gulati2020conformer}: an end-to-end ASR system based on Conformer. It has achieved the state-of-the-art on ASR, but may not perform well for noise-robust speech recognition.

\item \textbf{Cascaded SE-ASR}~\cite{subramanian2019speech}: a cascaded system consisting of a front-end SE module and a back-end ASR module. The system is optimized with ASR training objective only.

\item \textbf{Multi-Task Learning}~\cite{ma2021multitask}: a same structure as cascaded SE-ASR system, which adopts multi-task learning strategy to optimize the SE and ASR tasks simultaneously.

\item \textbf{Dynamic Weights}~\cite{groenendijk2021multi}: based on multi-task learning system, learns dynamic weights to balance SE and ASR training objectives, enabling their optimization with equal importance.

\item \textbf{PCGrad}~\cite{yu2020gradient}: based on multi-task learning system, employs PCGrad to avoid conflict between SE and ASR gradients.

\end{enumerate}

\begin{table}[t]
    \centering
    \vspace{-0.12cm}
    \caption{WER\% results of the proposed gradient remedy approach and competitive baselines on RATS Channel-A dataset.}
    \vspace{0.18cm}
    \label{table1}
    \resizebox{0.295\textwidth}{!}{
    \begin{tabular}{p{12em}|c}
        \toprule
        Method & WER\% \\
        \midrule
        E2E-ASR~\cite{gulati2020conformer} & 54.3 \\
        Cascaded SE-ASR~\cite{subramanian2019speech} & 53.1 \\
        Multi-Task Learning~\cite{ma2021multitask} & 51.8 \\
        Dynamic Weights~\cite{groenendijk2021multi} & 50.9 \\
        PCGrad~\cite{yu2020gradient} & 50.5 \\
        \midrule
        Gradient Remedy (ours) & \textbf{47.0} \\
        \bottomrule
    \end{tabular}}
    \vspace{-0.4cm}
\end{table}

\begin{table}[t]
    \centering
    \caption{WER\% results of the proposed gradient remedy approach and competitive baselines on CHiME-4 1-Channel Track dataset. ``Dev'' and ``Test'' denote the WER\% results on development set and test set, respectively. ``real'' and ``simu'' denote real noisy subset and simulated noisy subset, respectively.}
    \vspace{0.2cm}
    \label{table2}
    \resizebox{0.43\textwidth}{!}{
    \begin{tabular}{p{12em}|cc|cc}
        \toprule
        \multirow{2}{*}{Method} & \multicolumn{2}{c|}{Dev} & \multicolumn{2}{c}{Test} \\
        \cline{2-5}
         & real & simu & real & simu \\
        \midrule
        E2E-ASR~\cite{gulati2020conformer} & 8.1 & 9.6 & 14.9 & 16.1 \\
        Cascaded SE-ASR~\cite{subramanian2019speech} & 7.7 & 9.2 & 14.4 & 15.6 \\
        Multi-Task Learning~\cite{ma2021multitask} & 7.2 & 8.7 & 13.8 & 14.9 \\
        Dynamic Weights~\cite{groenendijk2021multi} & 7.0 & 8.6 & 13.5 & 14.6 \\
        PCGrad~\cite{yu2020gradient} & 6.9 & 8.4 & 13.3 & 14.5 \\
        \midrule
        Gradient Remedy (ours) & \textbf{6.3} & \textbf{7.8} & \textbf{12.2} & \textbf{13.4} \\
        \bottomrule
    \end{tabular}}
    \vspace{-0.35cm}
\end{table}

\vspace{-0.4cm}
\subsection{Results}
\label{ssec:results}
\vspace{-0.1cm}
We report experimental results in terms of word error rate (WER), as our target is ASR performance while SE is only auxiliary task.

\vspace{-0.3cm}
\subsubsection{Gradient Remedy vs. Other Competitive Methods}
\label{sssec:overall_compare}
\vspace{-0.15cm}
Table \ref{table1} summarizes the comparison between our proposed gradient remedy approach and other competitive methods on RATS Channel-A dataset. 
Specifically, E2E ASR system yields 54.3\% WER result, indicating the high difficulty of recognizing extremely noisy speech.
Cascaded SE-ASR system slightly improves the performance with help of SE module.
Multi-task learning method further lowers the WER result and achieves 2.5\% absolute improvement over E2E ASR baseline, indicating that the SE training objective is overall beneficial to downstream ASR task.
Dynamic weights and PCGrad strategies continue to improve by alleviating the gradient interference in multi-task learning, but they are quite limited.
Finally, our gradient remedy approach obtains the best result with 9.3\% relative WER reduction over multi-task learning baseline (51.8\%$\rightarrow$47.0\%), as well as 3.5\% absolute improvement over the best PCGrad baseline.

Table~\ref{table2} further compares our gradient remedy approach with the baselines on CHiME-4 1-Channel Track dataset.
We observe that the proposed GR approach achieves average relative WER improvement of 11.1\% over the multi-task learning baseline.

As a result, our proposed gradient remedy approach shows its effectiveness on both extremely noisy radio-communication RATS data and the normally noisy far-field CHiME-4 data.

\vspace{-0.27cm}
\subsubsection{Effect of Gradient Remedy on Interfered Gradients}
\label{sssec:effect_on_interfere}
\vspace{-0.1cm}

\begin{figure}[t]
  \centering
  \includegraphics[width=0.99\columnwidth]{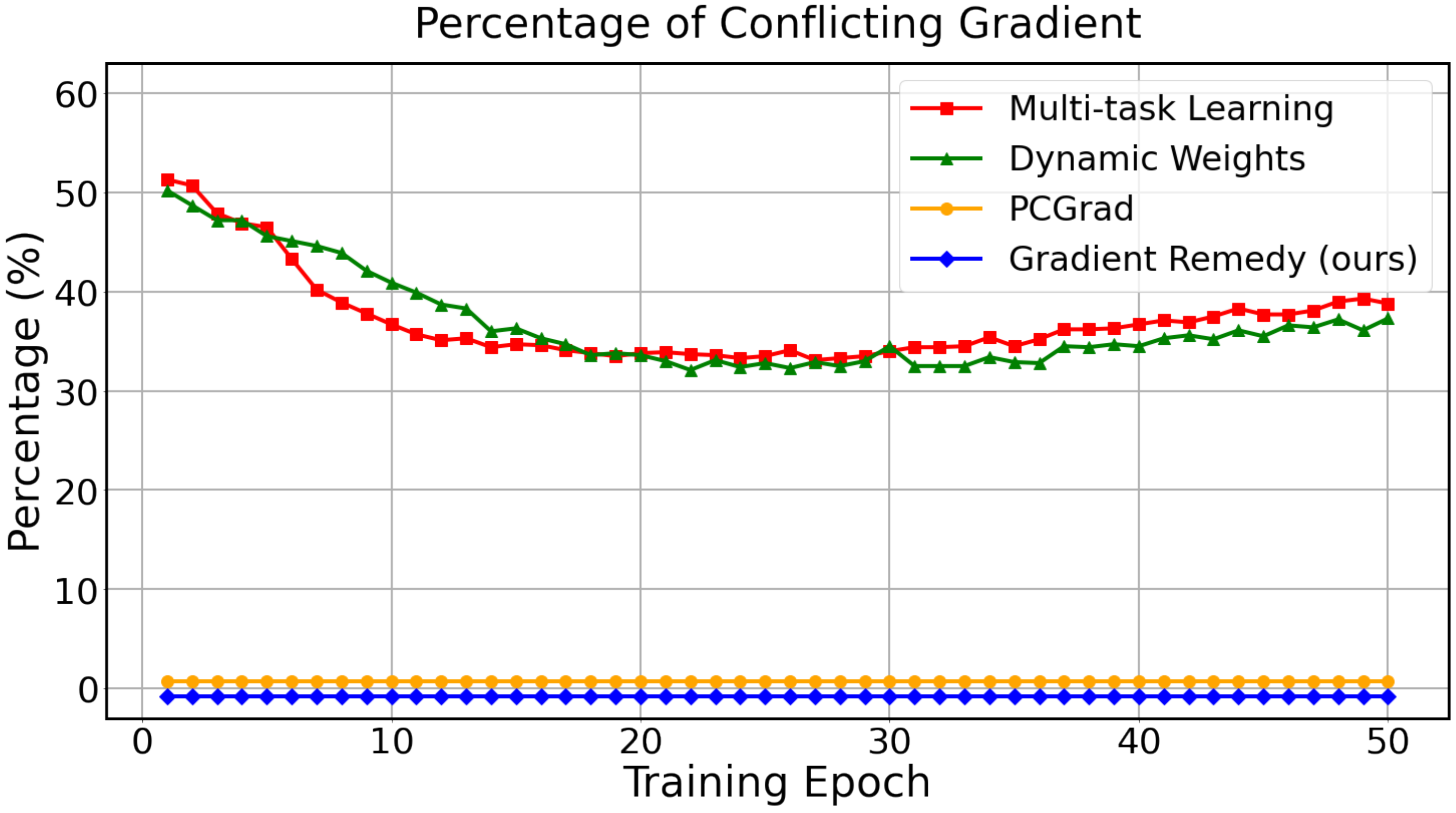}
  \vspace{-0.35cm}
  \caption{Percentage\% of conflicting gradient in all layers of SE module during training on RATS Channel-A dataset.
  The percentage value of each epoch is obtained by averaging all the batches in it.
  }\label{fig4}
  \vspace{-0.15cm}
\end{figure}

\begin{figure}[t]
  \centering
  \includegraphics[width=0.99\columnwidth]{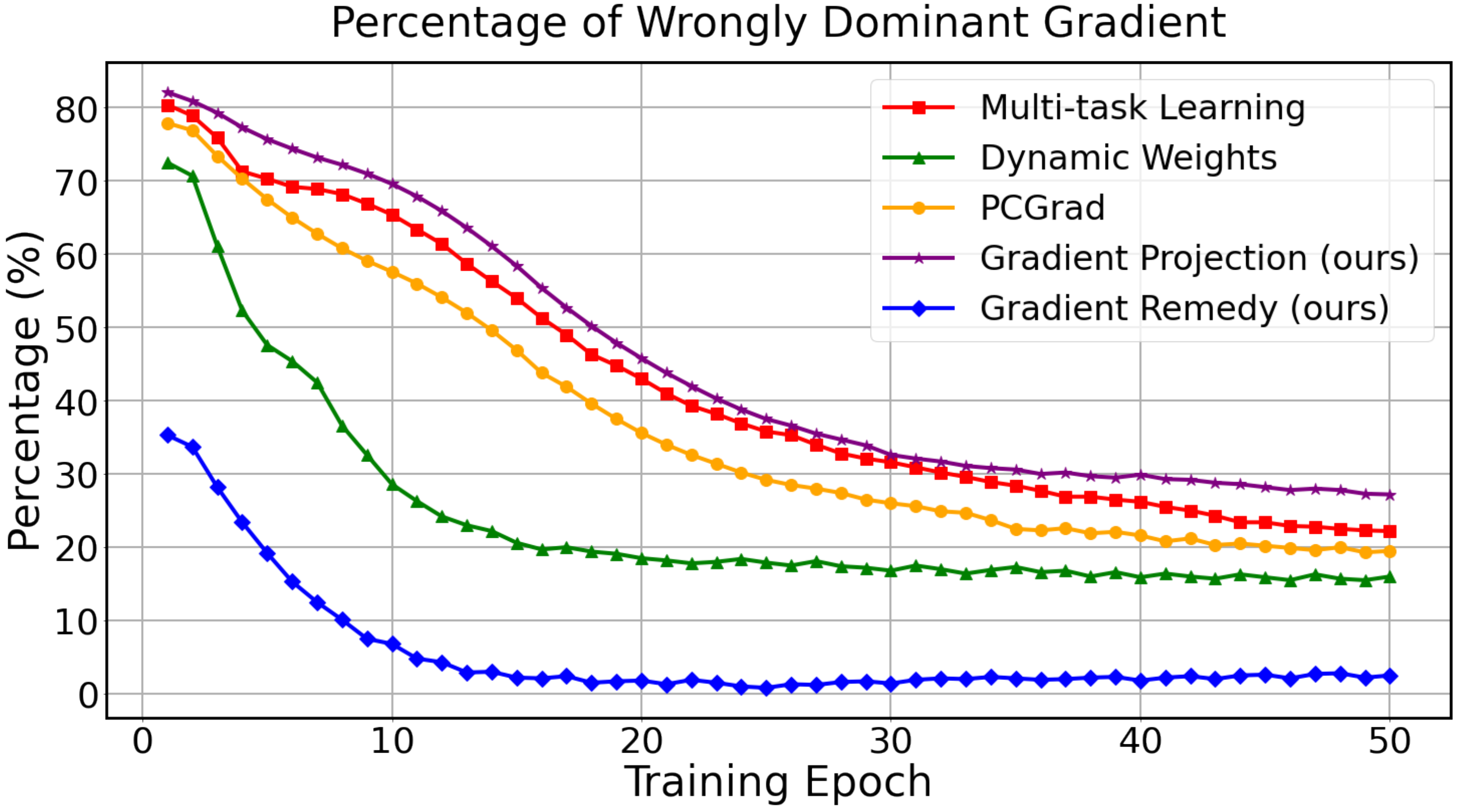}
  \vspace{-0.35cm}
  \caption{Percentage\% of wrongly dominant gradient ($K=5$) in all layers of SE module during training on RATS Channel-A dataset.}\label{fig5}
  \vspace{-0.48cm}
\end{figure}

To further show the effect of gradient remedy on interfered gradients, we present the percentage of conflicting and wrongly dominant gradients in all SE module layers in Figure~\ref{fig4} and \ref{fig5}.
Firstly, we observe that the multi-task learning system suffers a lot from gradient conflict, while the dynamic weights strategy cannot solve it.
In comparison, our gradient remedy approach completely removes the conflict, same as the PCGrad method.
Then in Figure~\ref{fig5}, we observe that the multi-task learning system also suffers seriously from wrongly dominant gradient.
PCGrad method slightly alleviates this problem as it reduces some of SE gradient's magnitude (See Figure~\ref{fig2}(a)), while our gradient projection method makes it worse, since the designed acute angle $\theta$ in Figure~\ref{fig2}(b) could lead to larger SE gradient magnitude than original.
Dynamic weights strategy further alleviates it by balancing different training objectives.
Finally, our gradient remedy approach best solves this problem by adaptively rescaling gradient magnitudes, with only few wrongly dominant gradients remaining.

\vspace{-0.27cm}
\subsubsection{Effect of Gradient Projection}
\label{sssec:effect_gp}
\vspace{-0.1cm}

\begin{table}[t]
    \centering
    \vspace{-0.2cm}
    \caption{WER\% results of gradient projection on RATS Channel-A dataset. To evaluate our designed dynamic $\theta$ in Equation~\ref{eq4}, we build several baselines with fixed $\theta$ throughout the training process.}
    \label{table3}
    \vspace{0.18cm}
    \resizebox{0.425\textwidth}{!}{
    \begin{tabular}{p{13em}|c|c}
        \toprule
        Method & $\theta^\circ$ & WER\% \\
        \midrule
        Multi-Task Learning~\cite{ma2021multitask} & - & 51.8 \\
        \midrule
        \multirow{6}{*}{\quad + Gradient Projection} & 90 & 50.5 \\
          & 60 & 50.0 \\
          & 45 & 49.6 \\
          & 36 & 49.5 \\
          & 30 & 49.8 \\
          & Dynamic (ours) & \textbf{48.8} \\
        \bottomrule
    \end{tabular}}
    \vspace{-0.4cm}
\end{table}

We further report the effect of gradient projection method in Table~\ref{table3}.
Based on the multi-task learning baseline, the PCGrad method with $\theta=90^\circ$ achieves some improvement (51.8\%$\rightarrow$50.5\%), indicating the effect of removing gradient conflict.
Then, our proposed gradient projection with fixed acute angle $\theta$ achieves further improvements, where fixing $\theta$ to $36^\circ$ performs the best (49.5\% WER), suggesting that it is effective to push SE gradient to assist in ASR optimization.
Finally, our designed dynamic $\theta$ in Equation~\ref{eq4} achieves the best performance with 3.0\% absolute WER improvement over multi-task learning baseline (51.8\%$\rightarrow$48.8\%), indicating the effectiveness of applying dynamic projection to steadily assist in ASR optimization.

\vspace{-0.25cm}
\subsubsection{Effect of Gradient Rescale}
\label{sssec:effect_gr}
\vspace{-0.1cm}

\begin{table}[t]
    \centering
    \caption{WER\% results of gradient rescale on RATS Channel-A dataset. 
    To evaluate our designed adaptive ratio $r=\cos\theta'$ for Equation~\ref{eq5}, we set $r=1/\sqrt{K}$ as well as constants ($\textit{i.e.}, 1/2, 1/3$) for comparison.
    Different threshold $K$ are also used for ablation study.}
    \label{table4}
    \vspace{0.18cm}
    \resizebox{0.48\textwidth}{!}{
    \begin{tabular}{p{9em}|c|>{\centering}p{2.4em}>{\centering}p{2.4em}>{\centering}p{2.4em}c}
    \toprule
    Method & \diagbox{$K$}{$r$} & $1/2$ & $1/3$ & $1/\sqrt{K}$ & $\cos\theta'$ \\
    \midrule
    Gradient Projection & - & \multicolumn{4}{c}{48.8} \\
    \midrule
    \multirow{5}{*}{\quad + Gradient Rescale} & $2$ & 49.6 & 49.9 & 49.4 & 49.1 \\
    & $3$ & 48.7 & 48.9 & 48.5 & 47.8 \\
    & $5$ & 47.7 & 48.0 & 47.6 & \textbf{47.0} \\
    & $6$ & 47.9 & 47.8 & 47.8 & 47.3 \\
    & $8$ & 48.5 & 48.1 & 48.3 & 47.6 \\
    \bottomrule
    \end{tabular}}
    \vspace{-0.47cm}
\end{table}

We finally report the effect of gradient rescale strategy in Table~\ref{table4}.
Firstly, our adaptive ratio $r=\cos\theta'$ consistently outperforms the $1/\sqrt{K}$ that simply rescales SE and ASR gradients to a similar magnitude, indicating the effectiveness of our adaptive rescale strategy.
In addition, constant ratios ($\textit{i.e.}, 1/2, 1/3$) can also improve WER but less than the above two designs.
Then for the choice of threshold $K$, we observe that small $K$ ($=2$) degrades the WER performance (48.8\%$\rightarrow$49.1\%), as too loose threshold would lead to overmuch rescale.
As $K$ increases, we achieve some improvements and obtain the best result at $K=5$, with 1.8\% absolute WER reduction over gradient projection baseline (48.8\%$\rightarrow$47.0\%).  
However, further increasing $K$ weakens the improvement, as too strict threshold significantly reduces the chance that our rescale strategy get triggered.
\vspace{-0.45cm}

\section{Conclusion}
\label{sec:conclusion}
\vspace{-0.2cm}
In this paper, we propose a gradient remedy approach to solve interference between task gradients in noise-robust speech recognition. 
Specifically, we first project the SE gradient onto a dynamic surface that is at acute angle to ASR gradient, in order to remove their conflict and assist in the ASR optimization.
Furthermore, we adaptively rescale the magnitude of two gradients to prevent the dominant ASR task being misled by SE gradient.
Experimental results on RATS Channel-A and CHiME-4 1-Channel Track datasets show that the proposed gradient remedy approach well resolves the gradient interference and significantly outperforms the competitive baselines.

\vfill\pagebreak

\ninept

\bibliographystyle{IEEEbib}

\end{document}